\documentclass[]{spie}  

\usepackage{amsmath,amsfonts,amssymb}
\usepackage{graphicx,caption}
\usepackage[colorlinks=true, allcolors=blue]{hyperref}
\usepackage{gensymb}

\title{Patterned liquid-crystal optics for broadband coronagraphy and wavefront sensing}

\author[a]{David S. Doelman}
\author[a]{Frans Snik}
\author[b]{Nathaniel Z. Warriner}
\author[b]{Michael J. Escuti}

\affil[a]{Leiden Observatory, Leiden University, P.O. Box 9513, 2300 RA Leiden, The Netherlands}
\affil[b]{Department of Electrical and Computer Engineering, North Carolina State University, Raleigh, NC 27695, USA}

\authorinfo{E-mail: doelman@strw.leidenuniv.nl}

\begin{document} 
\maketitle

\begin{abstract}
The direct-write technology for liquid-crystal patterns allows for manufacturing of extreme geometric phase patterned coronagraphs that are inherently broadband, e.g. the vector Apodizing Phase Plate (vAPP). We present on-sky data of a double-grating vAPP operating from 2-5 $\mu m$ with a 360-degree dark hole and a decreased leakage term of $\sim 10^{-4}$. We report a new liquid-crystal design used in a grating-vAPP for SCExAO that operates from 1-2.5$\mu m$. Furthermore, we present wavelength-selective vAPPs that work at specific wavelength ranges and transmit light unapodized at other wavelengths. Lastly, we present geometric phase patterns for advanced implementations of WFS (e.g. Zernike-type) that are enabled only by this liquid-crystal technology.
\end{abstract}


\section{INTRODUCTION}
\label{sec:intro}  

Direct imaging of exoplanets allows for detailed characterization by analyzing the separated light of the exoplanet from it's host star. This separation is realized using complex optical systems that selectively suppress light from the host star while maintaining high throughput for the companion \cite{Guyon2006TheoreticalCoronagraphs,Mawet2012ReviewSystems}. For ground-based coronagraphic high-contrast imaging instruments, reaching the contrast levels necessary to observe these faint companions is complicated by atmospheric turbulence and aberrations generated by the telescope itself \cite{Racine1999SpeckleCompanions}. Residual wavefront errors from imperfect correction by an extreme adaptive optics system as well as quasi-static optical aberrations lead to tip-tilt instabilities and a time-varying diffraction halo, degrading the performance of the coronagraph. \\
For each type of coronagraph, this degradation of the performance is different \cite{Guyon2006TheoreticalCoronagraphs,Krist2015NumericalPerformances}. Many different coronagraph designs have been proposed and they can be categorized as pupil plane coronagraphs, focal plane coronagraphs or a combination of both \cite{Ruane2015OptimizedApertures,Guyon2010HighMasks,Mawet2011ImprovedCoronagraph.}. In general, a coronagraph operating in the focal plane will suffer from stellar leakage generated by residual tip-tilt errors and partially resolved stars. Pupil plane coronagraphs are inherently insensitive to these aberrations. Pupil plane coronagraphs modify either the amplitude or the phase in the pupil plane to create destructive interference in the point spread function (PSF) at a region, here 'dark zone', where exoplanets are expected to be found. The apodizing phase plate (APP) coronagraph consists of a diamond turned optic from zinc selenide that introduces optical path differences with a differential height map, optimized to suppress star light from 2-9 $\lambda/D$ over a 180$\degree$ \cite{Codona2004ImagingBands, Codona2006ARadius,Kenworthy2007FirstPlate}. \\
The vector-apodizing phase plate (vAPP) is an upgraded version of the APP and induces geometric (or \\Pancharatnam-Berry) phase for circularly polarized light \cite{pancharatnam1956generalized,Berry1987TheLight,snik2012vector,Otten2014TheOutlook,Otten2017ON-SKYMagAO/Clio2}. The vAPP is a half-wave retarder where the fast-axis orientation changes as a function of position. The induced phase, $\phi$, is equal to plus/minus twice the fast-axis orientation $\theta$: $\phi = \pm 2\theta$, with opposite sign for the opposite circular polarization states. Geometric phase is by definition achromatic and the efficiency, the percentage of light that acquires this phase, depends on the retardance offset from half-wave. The broadband performance of the vAPP is therefore determined by the retardance as a function of wavelength.  \\
The vAPP is manufactured with liquid-crystal technology. A direct-write system is used to print the desired orientation (=phase) pattern in a liquid-crystal photo-alignment layer that has been deposited on a substrate \cite{Miskiewicz2014Direct-writingPatterns}. The induced orientation depends on the linear polarization of the incoming light. Multiple layers of self-aligning birefringent liquid-crystals are deposited on top with varying thickness and twist, generating the required half-wave retardance. \cite{Komanduri2012Multi-twistTransformation,Komanduri2013Multi-twistLayers}. The stack is a multi-layered twisted retarder (MTR). By tuning the twist and thickness of layers in the MTR, very high efficiencies can be achieved for multiple bands \cite{Otten2017ON-SKYMagAO/Clio2}.\\
The goal of this paper is to show new concepts and upgrades for the vAPP as well as applications of liquid-crystal technology for wavefront sensing purposes.

\section{The double-grating 360 vAPP}
The vAPP induces phase with opposite sign for circular polarization of opposite handedness. When unpolarized light comes in on a vAPP with a 180-degree dark zone, the flipped sign for each circular polarization state creates two PSFs where the dark zone is located on opposite sides and spatially separating them on the detector is necessary \cite{snik2012vector}. For 360-degree dark zones this is not the case as the dark zone is point symmetric. However, non-perfect half-wave retardance of the vAPP introduces a leakage PSF that is also imaged on top of the two PSFs, degrading the contrast for small inner working angles (IWAs). Previously, a quarter-wave retarder and a Wollaston prism were used to physically  separate the PSFs, in addition to reducing the leakage term by a factor two. For small IWAs a leakage of 1\% introduces light on the order of $10^{-4}$ at these locations, limiting the contrast that can be obtained.\\ 
Both problems can be solved by adding a polarization grating (PG) to the pattern \cite{Otten2014TheOutlook}. The PG is a simple phase ramp in geometric phase such that both PSFs now acquire opposite tilt and are separated on the detector. In addition, the leakage-term does not acquire tilt and is imaged on the optical axis, also separated from both coronagraphic PSFs. 
   \begin{figure} [ht]
   \begin{center}
   \begin{tabular}{c} 
   \includegraphics[height=8cm, trim={1cm 3cm 0cm 3.5cm},clip]{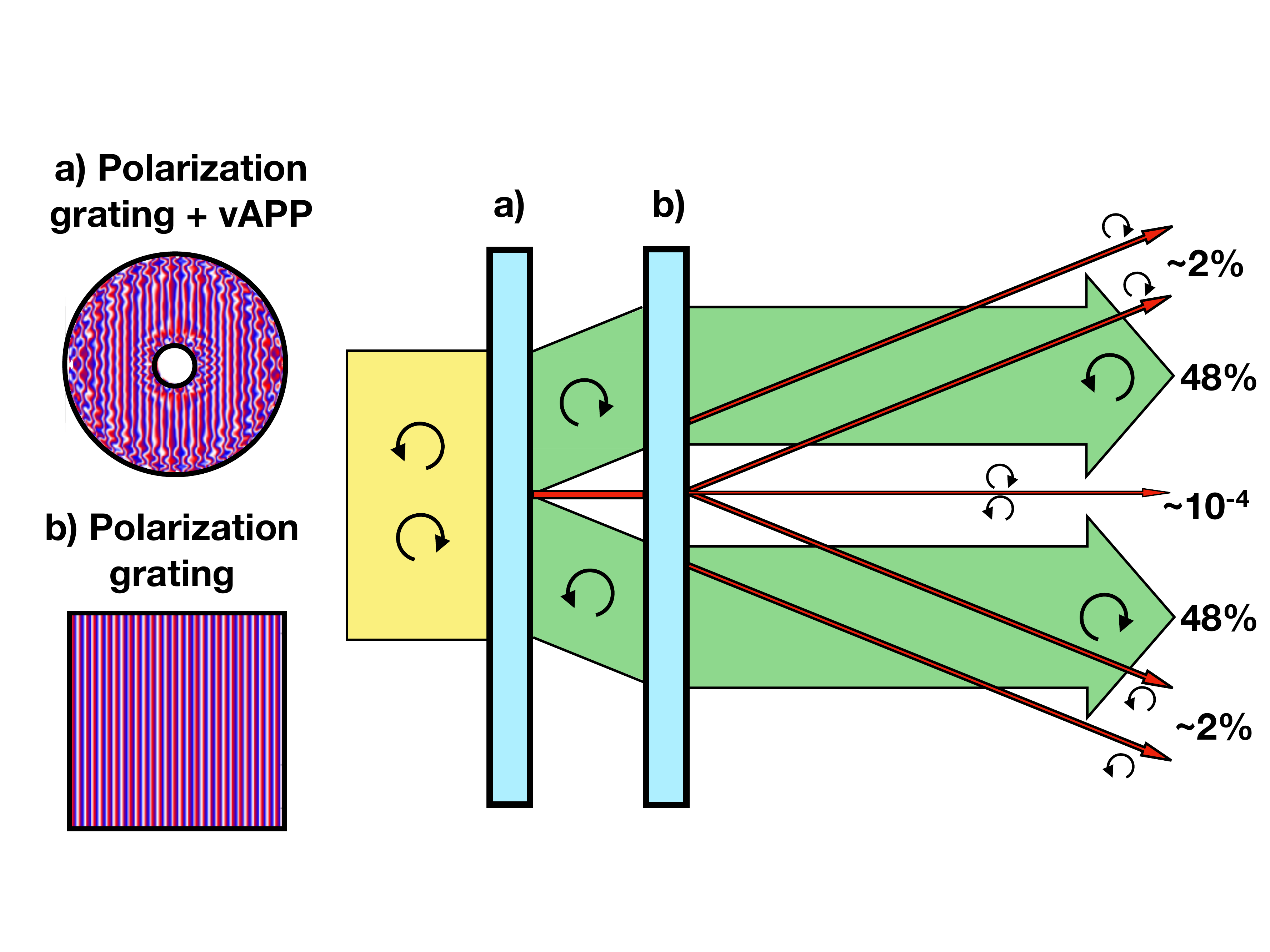}
   \end{tabular}
   \end{center}
   \caption[example] 
   { \label{fig:doublegrating} 
   Schematic diagram of the double-grating. Unpolarized light (yellow) enters the grating-vAPP. The grating-vAPP consists of a two patterns added together: a polarization grating and a vAPP that generates a 360-degree dark zone. The combined pattern splits the two circular polarization states and applies the phase that generates the dark zone (green). The second polarization grating reverses the tilt from the first polarization grating pattern and the apodized light continues without tilt. A leakage term (red) goes through the grating-vAPP  without acquiring any phase. This leakage term is apodized by the second grating and is diffracted. Any offset from half-wave retardance for the second polarization grating generates three leakage terms. The first is a double leakage term, that only contains 0.01\% of the light and the other two are diffracted beams of the grating-vAPP. Note that the latter leakage terms have been apodized by the first grating-vAPP and have a dark zone.}
   \end{figure}
The downside to this trick is the wavelength-dependent dispersion that is introduced for the coronagraphic PFSs.
 The consequence is that it is not possible to use a fiber injection unit for high resolution spectroscopy of the exoplanet light.  \\
We introduce the double-grating vAPP to reduce the leakage terms and make use of the point-symmetry in the system, as shown in Fig. \ref{fig:doublegrating}. The grating-vAPP and the second grating are in the same pupil plane and can be as close as a few mm \cite{Escuti:16}. The second grating effectively cancels the grating on the grating-vAPP, such that the residual is a standard vAPP with a 360-degree dark zone. Note that polarization gratings have a "memory" for the handedness of circular polarization and that is why the double-grating can operate with close to 100\% efficiency. The two gratings need to have the same grating frequency and orientation because the change in circular polarization state from the first (half-wave) grating-vAPP automatically changes the sign necessary to cancel the phase ramp with the second PG. Therefore, precise alignment of both gratings is key. \\
The real advantage of such a system becomes clear if we look at the leakage term of the first vAPP. This leakage term is apodized by the PG and is scattered to high spatial frequencies. Only the leakage term of the PG from the leakage light of the grating-vAPP ends up at the location of the coronagraphic PSFs. For 1\% leakage per MTR, the total leakage is reduced to $10^{-4}$ and for a clear aperture, the light that is introduced in the dark zone at 2 $\lambda/D$ is on the order of $10^{-6}$. Furthermore, the light of the exoplanet is no longer dispersed and can be fed into a fiber for spectroscopic analysis. \\
The vector-vortex coronagraph could also benefit from the double-grating concept. A double-grating vector-vortex is created by replacing the vAPP pattern by a vortex pattern and placing the device in the focal plane. It is an elegant solution for a broadband vortex coronagraph without the need to filter the leakage light.  

\section{The double-grating vAPP for LMIRCAM on the LBT}
\begin{figure}
\centering
\begin{minipage}{\textwidth}
  \centering
   \captionsetup{width=.9\linewidth}
  \includegraphics[width=\linewidth, height = 8cm]{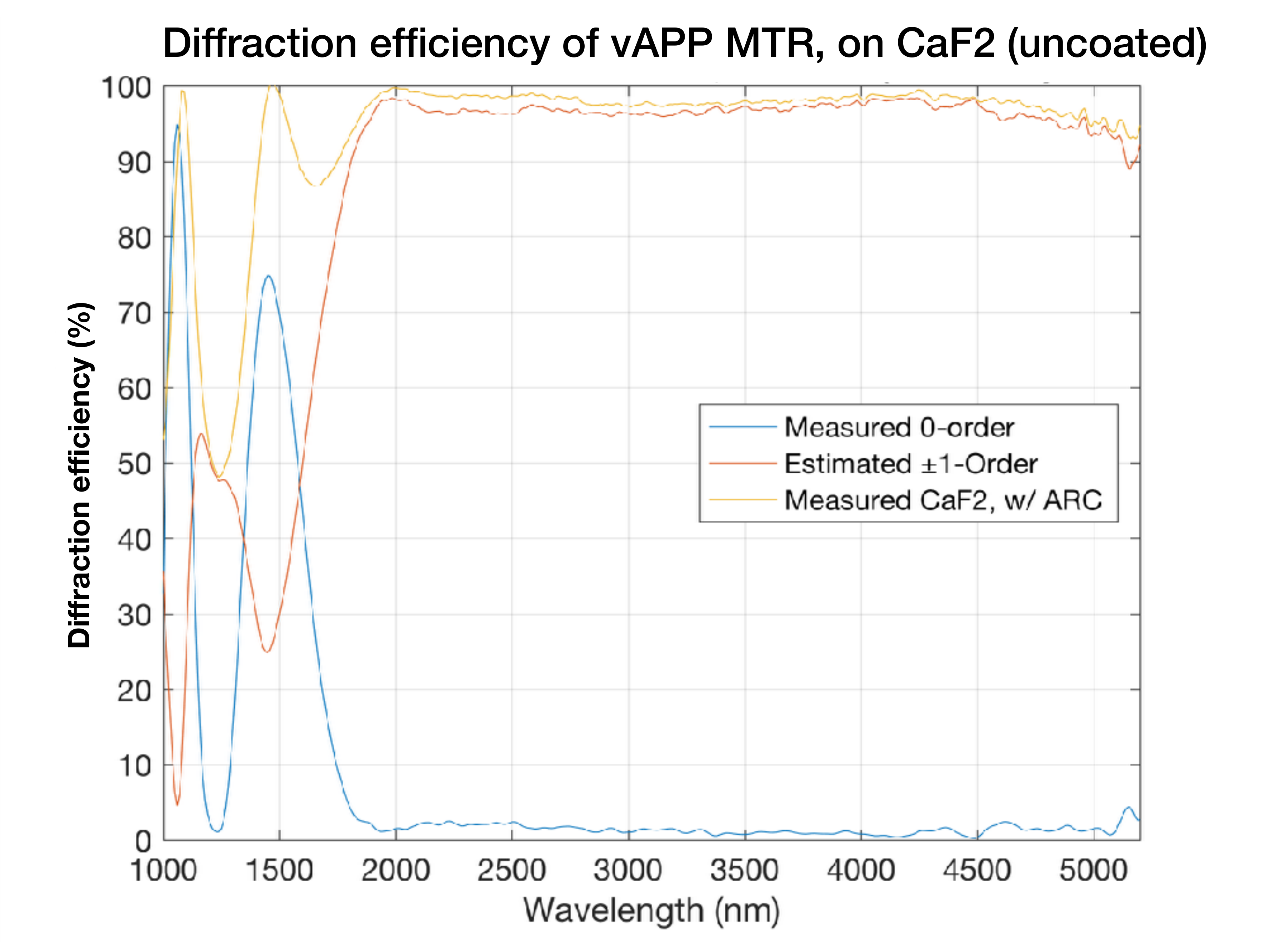}
  \caption{\textit{Blue:} 0-order diffraction efficiency (leakage) of the MTR used for the double-grating vAPP. The leakage is less than 3\% for 2-5 micron. \textit{Yellow:} Transmission of the CaF2 substrate measured with a Fourier transform spectograph. \textit{Orange:} Estimated $\pm$first order diffraction efficiency, calculated from the blue and yellow curves.}
  \label{fig:diffeff}
\end{minipage}%
\end{figure}

\begin{figure}[ht]
\centering
\begin{minipage}{.65\textwidth}
  \centering
   \captionsetup{width=.95\linewidth}
  \includegraphics[width=\linewidth, trim={2cm 0cm 0cm 0cm},clip]{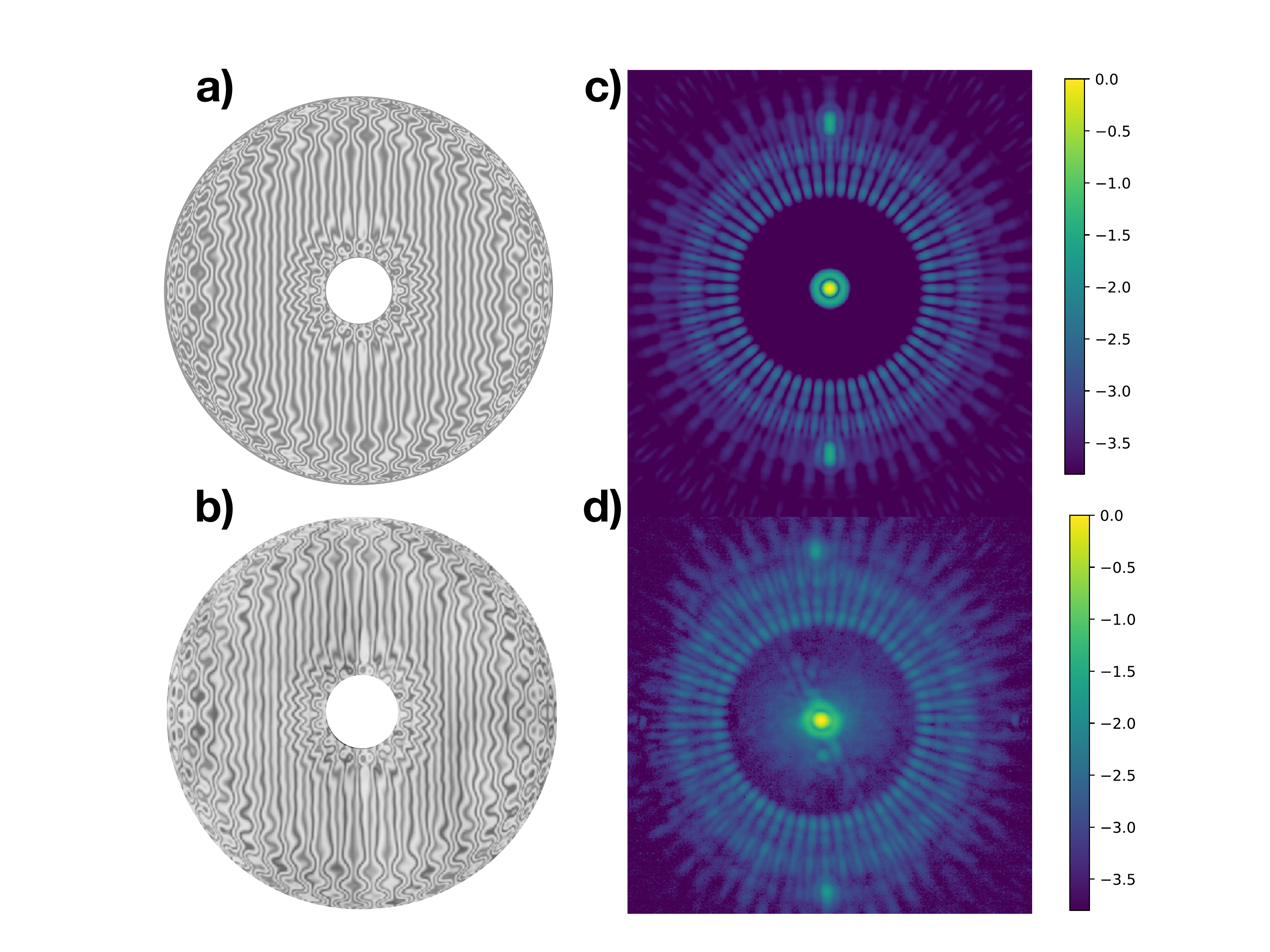}
  
\end{minipage}%
\begin{minipage}{.35\textwidth}
  \centering
  \captionsetup{width=.95\linewidth}
\caption{
\textbf{a)} Simulated monochromatic transmission through the grating-vAPP between polarizers. \textbf{b)} Measured broadband transmission of the grating-vAPP between polarizers. The manufactured grating-vAPP shows close to no deviation from the simulated pattern. \textbf{c)} Simulated PSF of the double-grating vAPP without aberrations for 3.6-3.9 micron. \textbf{d)} Measured PSF of the double-grating vAPP at LMIRCAM in L-band under bad seeing conditions. The overall structure of the measured PSF closely resembles the theoretical PSF. The residual halo of the atmospheric speckles limit the obtained contrast. \label{fig:phase_PSF}}

\end{minipage}
\end{figure}

A double-grating vAPP was installed on the LMIRCAM instrument \cite{skrutskie2010large} at the large binocular telescope (LBT) in September 2016. Here we will briefly go into the design, a more thorough description of the manufacturing process can be found in Otten et al. 2017 \cite{Otten2017ON-SKYMagAO/Clio2}. The phase pattern was optimized to create a 360-degree dark zone from 3-12 $\lambda/D$. The optimization was done with simulated annealing using 20 radial polynomial modes and a sinusoidal variation in the azimuthal direction with 28 periods in 2$\pi$ radians. To guarantee manufacturability, the azimuthal frequency was changed to 14 periods within a normalized radius of 0.3. The Strehl ratio is 37$\%$.\\  The double-grating vAPP consists of three 1 inch CaF2 substrates glued together to form one optic. Both telescopes can be used simultaneously with the double-grating vAPP as the double-grating patterns have been printed twice, one for each pupil. The first substrate has a chromium amplitude mask with ND $>4$ that defines both pupils of the optic. The second substrate has a 3-layered MTR (3TR) with two grating-vAPP pattern that align with the pupils. The third substrate has a 1-degree wedge and has a 3TR with the PG pattern. Optimization of the 3TR was done to have a high diffraction efficiency from 2-5 micron. The final diffraction efficiency can be seen in Fig. \ref{fig:diffeff}, where the 0-order leakage is less than 3\% from 2-5 micron.  Therefore the leakage PSF will have at most an intensity of $10^{-5}$ in the dark zone and is not limiting the performance of the coronagraph. \\
Before the double-grating vAPP was delivered to the LBT, the fast-axis orientation pattern of the grating-vAPP was investigated with a microscope. Microscopic images of the grating-vAPP between crossed polarizers stitched together can be seen in Fig. \ref{fig:phase_PSF} b). A simulation of the same pattern between polarizers can be seen in panel a) of the same figure. The fact that the manufactured pattern closely resembles the simulated pattern implies that the manufacturing was be done with great precision. \\
After installation, the first on-sky data was taken at the 12th and 13th of January 2017 where Regulus was observed with and without double-grating vAPP in L-band. Beam-switching was done every 50 frames with with an individual integration time of 0.29 sec. In total, 289 images were taken with the double-grating vAPP for a total integration time of 84 seconds. The data reduction was done by subtracting the median of the images with a different nod-direction from each individual frame. This was followed by removing additional background by subtracting the median of each column for every pixel in that column in the image. Because of poor seeing conditions ($\sim$2 arcsec), the frames with the highest Strehl were selected and sub-pixel aligned. The final image, shown in Fig. \ref{fig:phase_PSF} d) is the median combination of the best 28 frames. A simulated PSF without any wavefront aberrations is shown in panel c) for comparison. 
The PSF was simulated for 3.6-3.9 micron while the data was taken in L-band, 3.4-3.95 micron. An absorption feature of the liquid-crystals and the glue that was used around 3.4 micron degrade the transmission significantly between 3.4-3.6 micron. The overall shape of the PSF matches very well with the simulated PSF, proving that the concept of the double-grating works on-sky. However, the poor seeing conditions significantly degrade the   vAPP and limit the obtained contrast. \\
The double-grating vAPP can be used in combination with the ALES instrument, an AO corrected integral field spectrograph operating from 2-5$\mu$m \cite{Skemer2015FirstLBT}. With 360-degree search space from 3-12 $\lambda/D$ with $10^{-5}$ contrast operating over the full wavelength range of ALES, the combination provides a tool for simultaneous multi-band exoplanet observations. 

\section{The grating mask}
\begin{figure} [ht]
   \begin{center}
   \begin{tabular}{c} 
   \includegraphics[width = 0.8\textwidth, trim={0cm 5.5cm 0cm 5.5cm},clip]{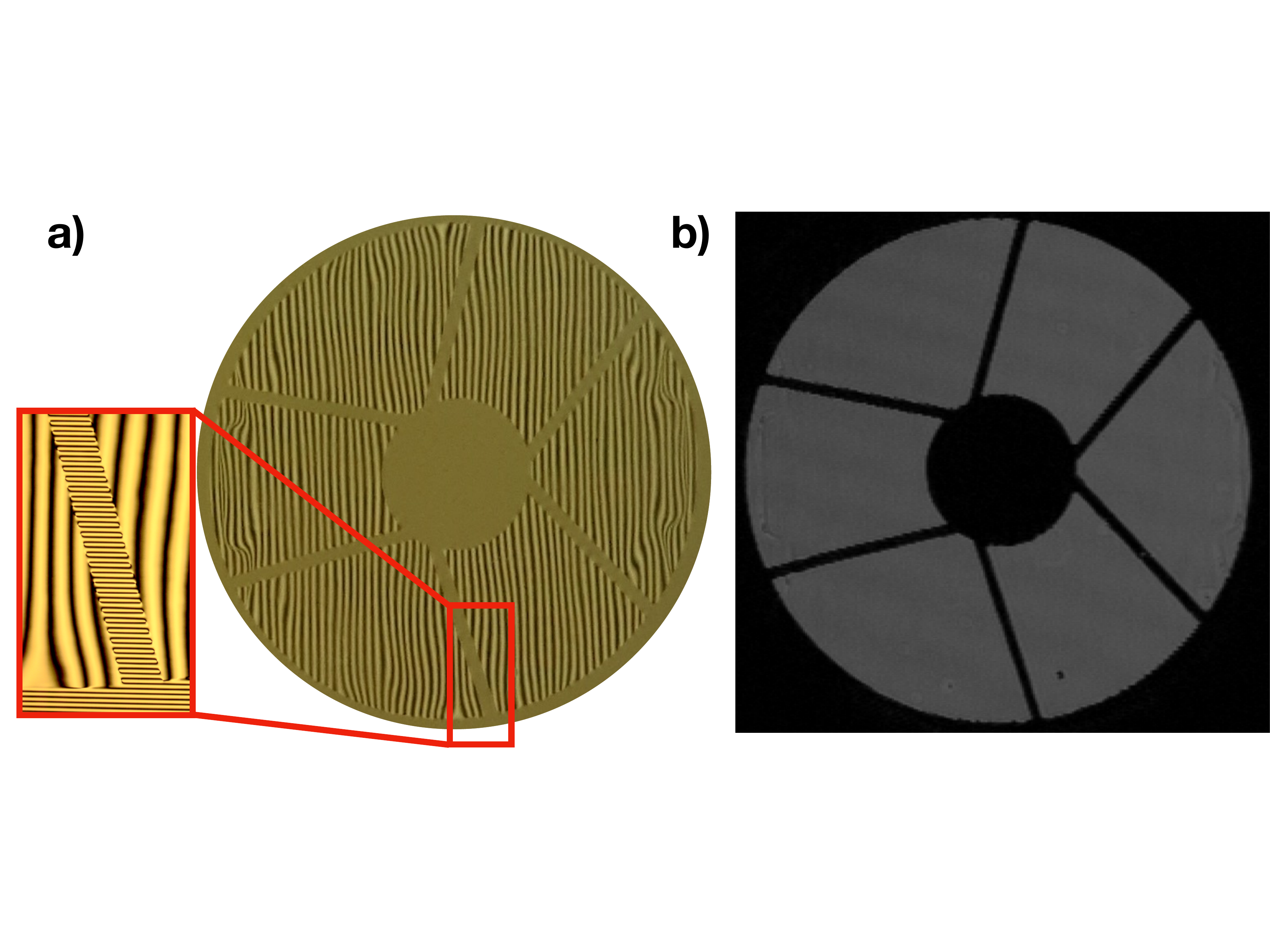}
   \end{tabular}
   \end{center}
   \caption[example] 
   { \label{fig:imagegm}
   \textbf{a)} Photograph of the vAPP with a simplified WFIRST pupil between crossed polarizers, manufactured for testing the grating mask. The zoomed in microscope image shows the high-frequency grating used for the grating mask that is not captured by the normal camera. \textbf{b)} Reimaged pupil with a field stop inserted in the focal plane. The field stop removes the diffracted light from outside the pupil defined by the grating mask. Even spiders with a width of a few times the grating period are removed extremely well. 
   }
   \end{figure}

A polarization grating (PG) applies a geometric phase tilt to the incoming electric field. The tilt is applied locally unlike a regular grating \cite{Escuti:16}.  The PG added to a normal vAPP, the grating-vAPP, was introduced in Otten et al. 2014 \cite{Otten2014TheOutlook}. The combination is used to separate both coronagraphic PSFs from each other and the leakage term. Previous grating-vAPPs added a PG both inside and outside the pupil defined by the amplitude mask. For the double-grating vAPP installed on the LBT, the grating is only applied to the pupil defined by the phase pattern and is padded with zero phase outside. Any misalignment in the pupil mask causes light with zero phase to enter the second PG and is diffracted to the leakage terms. Therefore, if the diameter of the chromium amplitude mask is somewhat bigger than the vAPP phase pattern, the effective pupil diameter set by the vAPP is not changed. The coronagraphic PSF is not altered except for more light in the leakage term. \\
   \begin{figure} [ht]
   \begin{center}
   \begin{tabular}{c} 
   \includegraphics[height=10cm]{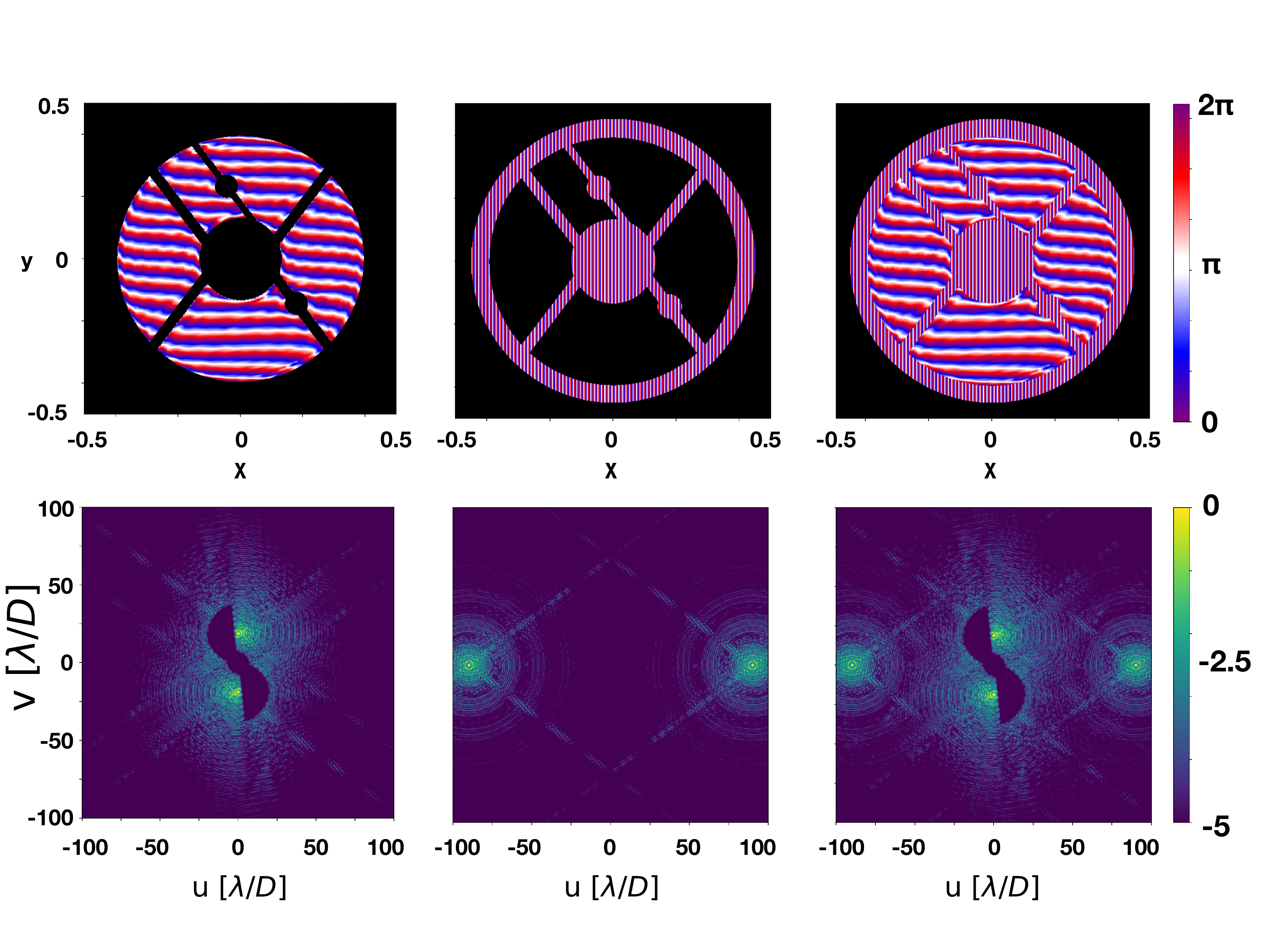}
   \end{tabular}
   \end{center}
   \caption[example] 
   { \label{fig:gratingmask}
   \textit{Top left:} Phase of a grating-vAPP for an undersized pupil of the Subaru telescope. An amplitude mask for the pupil is shown in black. \textit{Bottom left:} Logarithmic plot of the PSF of the grating-vAPP, showing two coronagraphic PSFs for opposite handedness of circularly polarized light. The grating-vAPP was simulated to be a perfect half-wave retarder, so no leakage is present. \textit{Center:} The grating mask (top) and resulting PSF (bottom) that will be used as an amplitude mask. The grating has 90 periods across the diameter of the pupil. \textit{Right:} The phase and PSF of the grating-vAPP combined with the grating mask. The incoming light has a pupil of radius 0.45 (black) and the light is separated in a grating mask component and a grating-vAPP component. A small misalignment of the pupil does not affect the coronagraphic PSF, relaxing alignment tolerances. }
   \end{figure}
Alignment tolerances of the amplitude mask can be significantly relaxed by this simple trick and it is possible because the Fourier transform is a linear operation. A slightly different implementation of the PG allows for a standard grating-vAPP to get rid of an amplitude mask entirely. Applying a PG outside the pupil with high spatial frequency in a different direction than the PG in the grating-vAPP, scatters light outside the pupil far away from the coronagraphic PSFs in the focal plane. This can be seen in Fig. \ref{fig:gratingmask}. A simple spatial filter in an intermediate focal plane can remove the light outside the pupil. We call this concept a grating mask. Lab testing of the grating mask was done for a vAPP with a simplified WFIRST pupil. The grating mask has 8 pixels of 8.75 micron per period for a 1cm pupil. The simplified WFIRST pupil was chosen for the thin spiders and complex mask design. An image between crossed polarizers of the manufactured design of the vAPP is shown on the left in Fig. \ref{fig:imagegm}. The microscopic image of the vAPP shows a spider with the grating pattern. Only a few periods fit in the spider, however this does not affect the performance of the grating mask as shown on the right. Panel b) is the reimaged pupil with a field stop added to the focal plane. The light outside the simplified WFIRST pupil is therefore blocked and does not appear in the reimaged pupil. \\
When applied in a real telescope, there are two things to consider. First, one has to be careful where the diffracted light from outside the pupil ends up. If not blocked adequately, the light might be scattered back onto the CCD by reflections inside the instrument. The direction of the PG is easily controlled, so prior knowledge of a preferred direction and a field stop can solve this problem. 
Second, for a deviation in retardance from half-wave, a few percent of the light that would be normally blocked by an amplitude mask would end up on the science camera. This effect will be small when a grating-vAPP with a high spatial frequency is manufactured, as the coronagraphic PSFs are have enough spatial separation from the leakage term that it can be ignored. In addition, the grating mask will only be used to relax alignment tolerances. The pupil defined by the grating mask will be undersized a few \% compared to the telescope pupil. Overall only a very small increase in leakage is expected.  

\section{Near-infrared vector-APP in SCExAO feeding CHARIS}
\label{sec:SCExAO}
Finding an exoplanet with direct imaging using a coronagraph is only the first step in the characterization process. Spectroscopic exoplanet characterization is necessary to constrain properties such as temperature, composition, presence of atmospheres and ultimately habitability. 
\begin{figure} [ht]
\begin{center}
\begin{tabular}{c} 
\includegraphics[width = 0.9\linewidth, trim={0cm 1cm 0cm 2cm},clip]{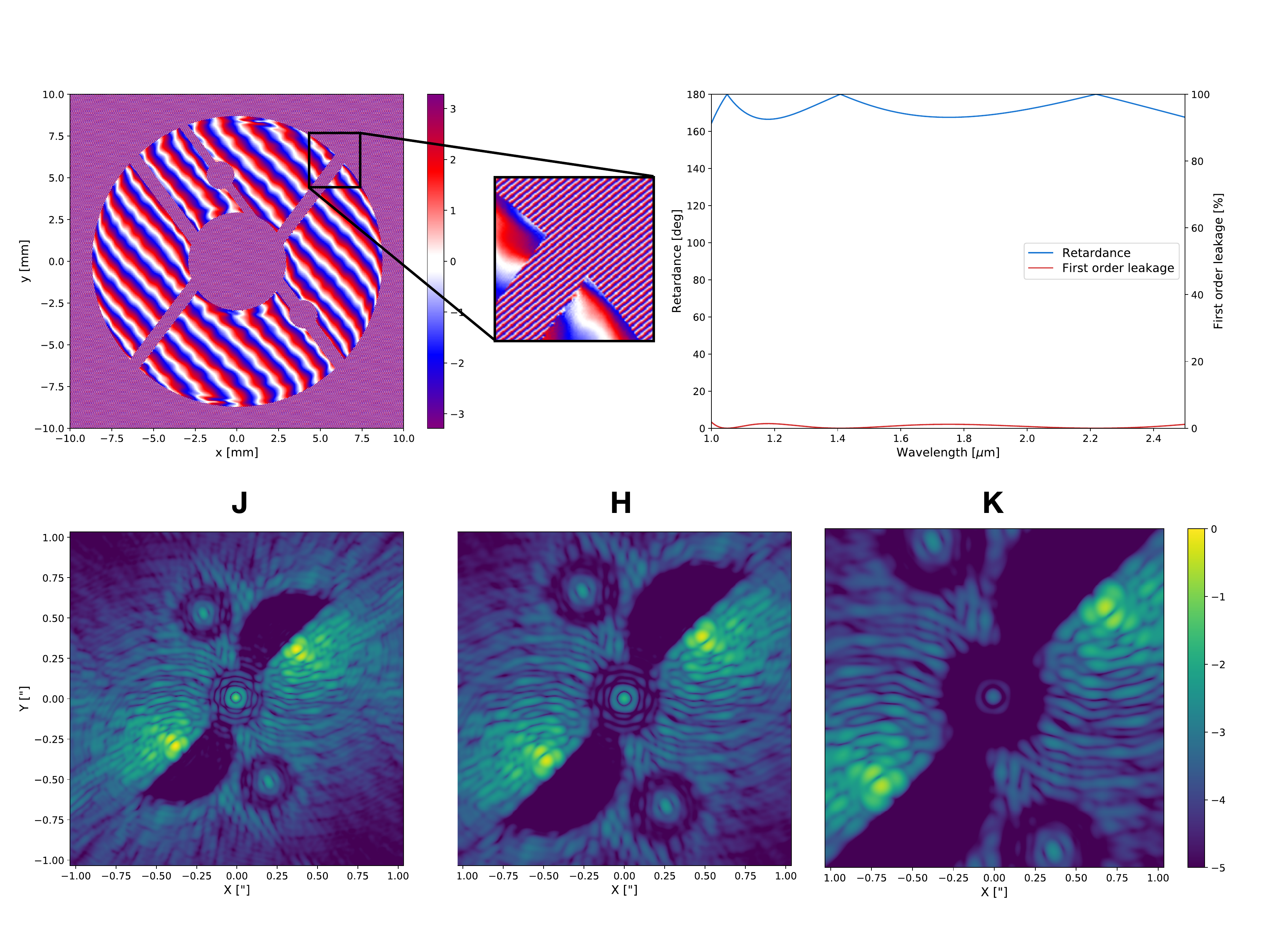}
\end{tabular}
\end{center}
\caption[example] 
{\label{fig:SCExAOdesign} \textit{Top left:} Design of the grating-vAPP phase pattern for SCExAO in radians. The pupil has been undersized by 1.5\% and is combined with a grating mask for alignment tolerances. \textit{Top right:} Properties of the liquid-crystal recipe used for the grating-vAPP. The retardance (blue) is very close to half-wave from J-K, resulting in a leakage (red) that is less than $2\%$ for $100\%$ bandwidth. \textit{Bottom:} Simulated PSFs of the grating-vAPP on the FOV of the CHARIS instrument for R=15 in the bands J, H and K. The grating-vAPP produces two D-shaped dark zones with in inner working angle of $2\lambda/D$ and an outer working angle of $11\lambda/D$. The central PSF is the leakage term. The two spots north and south of the leakage term are phase diversity holograms with a defocus of one fifth of a wave peak to valley. The holograms have opposite defocus and can be used for low-order focal plane wavefront sensing.}
\end{figure}
A spectrograph with fiber injection or an integral field unit (IFU) can be combined with the coronagraph to obtain a planetary spectrum. In recent years the first spectra of directly imaged planets have been observed with SPHERE and GPI \cite{Konopacky2013DetectionAtmosphere.,Ingraham2014GEMINID,Rajan2017CharacterizingExoplanet}. For precise characterization, observations across multiple bands are necessary. Coronagraphs that allow to search for planets close to their host star ($<$ 1 arcsec separation) are almost all chromatic and need to be manufactured for each band separately. For an IFU like CHARIS at the Subaru telescope, this is inefficient use of the instrument. CHARIS \cite{Peters-Limbach2013TheTelescope,Brandt2014CHARISInstrumentation}  can observe from 0.9-2.4 $\mu$m (y-K) simultaneously, so only a coronagraph with 100\% bandwidth can use the full potential of such an instrument. \\
Here we present a new grating vAPP coronagraph that will be placed in the SCExAO instrument\cite{Martinache2009TheProject,Jovanovic2013SCExAOInstrument,Jovanovic2016TheScience} at the beginning of August 2017 to feed the CHARIS instrument. 
The phase pattern of the grating-vAPP can be seen in the top left of Fig. \ref{fig:SCExAOdesign}. The grating of the grating-vAPP is oriented at 39 degrees and has 15 periods across the pupil to fit the dark zone on the detector for K-band. The grating mask with 252 periods across the pupil is oriented at -51 degrees to minimize light scattering in the dark zones. The incoming pupil is 1764mm and the grating-mask is undersized by 1.5\% to allow for 130 micron alignment tolerances in either direction. \\
A new liquid-crystal recipe is optimized for J, H and K band with a 98\% efficiency. The retardance and first order leakage are shown on the top right of Fig. \ref{fig:SCExAOdesign}. Because the leakage terms are smallrrd, the dark zones can be placed close to the leakage PSF, allowing for a compact design. This is necessary for the 2.07" x 2.07" FOV of the CHARIS instrument. The simulated PSF for R=15 for J, H and K is shown in the bottom of Fig. \ref{fig:SCExAOdesign}. The D-shaped dark zones have a contrast of $10^{-5}$ from $2\lambda/D$ to $11\lambda/D$ and are located on different sides of the star, giving close to 360-degree coverage. \\
Phase diversity holograms with 1.3 radians peak to valley defocus have been added. The spots are similar to Wilby et al. 2016 \cite{Wilby2016TheEnvironments}. Both spots have opposite handedness of circular polarization and therefore differ in sign of the defocus aberration. They can be used for classical phase diversity \cite{gonsalves1979wavefront,gonsalves1982phase,Paxman1988OpticalDiversity} to measure residual low-order aberrations in the focal plane, making them ideal for non-common path aberrations in the system. A dark zone was added underneath the holograms to improve the signal to noise ratio. The same was done for the leakage term, that now can be used for photometric and astrometric referencing. 
\section{Wavelength selective multi-twist retarders}
A great advantage of the liquid-crystal technology is that the fast-axis orientation and retardance of the devices are decoupled. One can optimize the retardance for a different wavelength range while using the same orientation pattern and the device would create the same PSF at the new wavelength range. The liquid-crystal recipe however, does not have to be optimized to be as close to half-wave as possible, this only determines its efficiency at those wavelengths. By changing the twist and thickness of the layers in the MTR, one can in principle make the retardance follow any continuous curve as function of wavelength, given a large amount of layers \cite{Hornburg2014MultibandRetarders}. For the coronagraphic application of this technology, the retardance could switch between 0 and 180 degrees for different wavelength ranges \cite{Kenworthy2016High-contrastMETIS}. A retardance of zero means that the efficiency is zero and no geometric phase is acquired by the light traveling through the vAPP. The vAPP is effectively switched off at those wavelengths, while it is switched on in the range where the retardance is close to 180 degrees. We call this wavelength-selective coronagraphs. \\
In a high-contrast imaging system, the wavefront sensor arm and science arm are battling for photons. Having a good wavefront correction needs to be balanced with integration times of the science camera. Going off-band with wavefront sensing is not ideal as aberrations depend on wavelength. Operating the wavefront sensor as close as possible to the science wavelengths increases the performance of the AO system in the bands where the coronagraph operates. The first advantage of the wavelength-selective coronagraph is that it can be tuned to very specific windows.
\begin{figure} [ht]
\begin{center}
\begin{tabular}{c} 
\includegraphics[height=10cm,width = 0.9\linewidth]{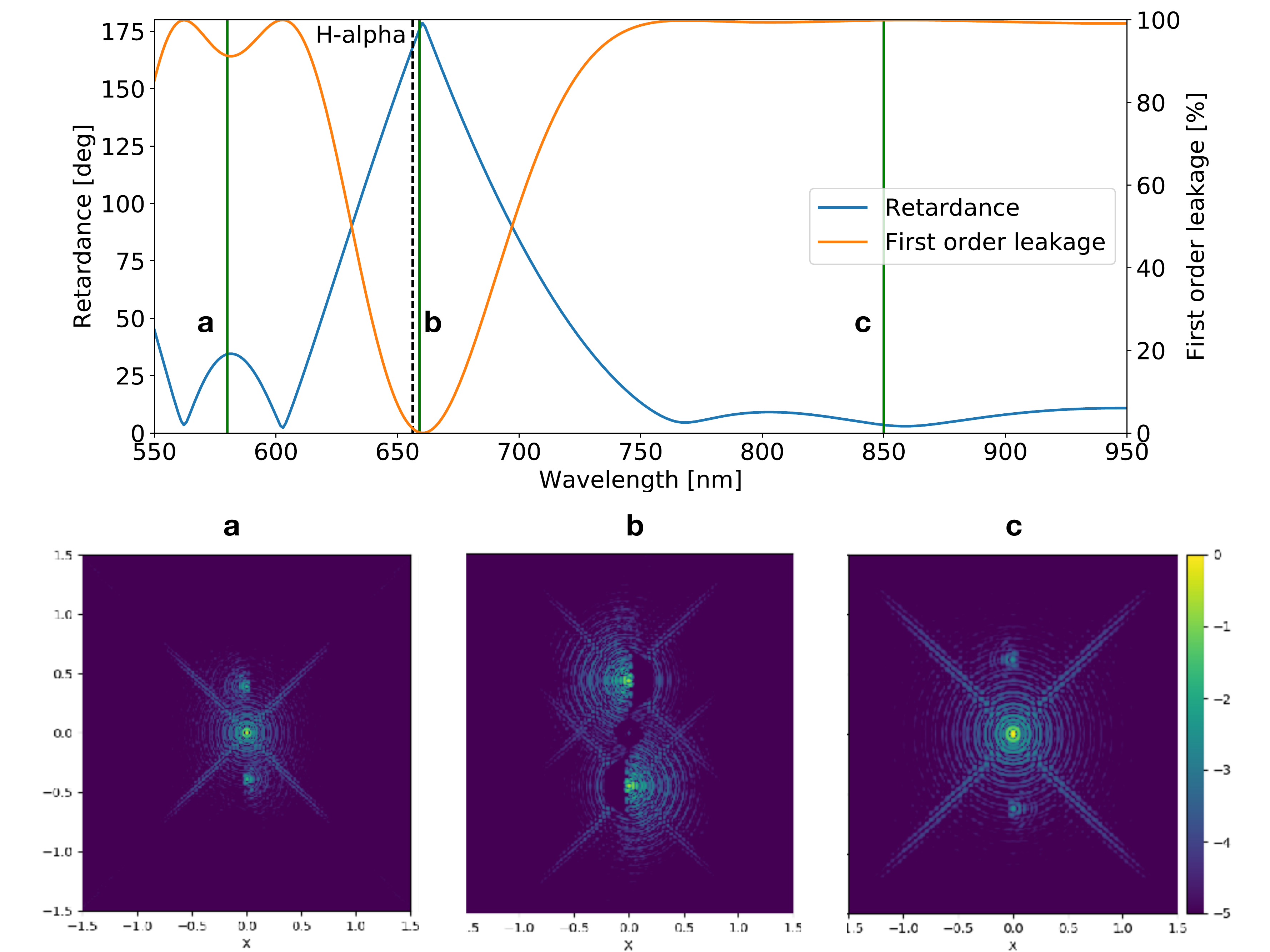}
\end{tabular}
\end{center}
\caption[example] 

{
\textit{Top:} Simulated retardance (blue) and first order leakage (orange) as function of wavelength for a 3-layered MTR. The total thickness is 20.7 micron and it was optimized to have 0 retardance everywhere except for a 10 nm band around H$\alpha$ wavelength (656.28 nm), where the retardance is close to 180 degrees. \textit{Bottom:} The simulated PSFs of a grating-vAPP with the retardance profile shown above. The green lines in the top figure correspond to the wavelengths the bottom figures were generated. Left of H$\alpha$ the leakage term dominates and almost no coronagraphic PSF can be seen (a). This is the same for wavelengths larger than 750nm (c). Around H$\alpha$ the leakage term disappears and the grating-vAPP operates as normal (b).\label{fig:wavelength-selective}
}
\end{figure}
An example of this is shown in Fig. \ref{fig:wavelength-selective}. The retardance is close to 180 degrees for 656nm, H$\alpha$, close to zero for smaller than 600nm and larger than 750nm. The light outside the H$\alpha$ band can be used by a wavefront-sensor, increasing the signal-to-noise ratio. Instruments like MagAO-X\cite{Males2016TheMagAO-X} can benefit from the increased flux to reach very high Strehl ratios at the H$\alpha$ wavelength.  \\
The second advantage of the wavelength-selective coronagraph is that the wavefront-sensor can be placed \textbf{after} the coronagraph. The simulated PSFs shown in Fig. \ref{fig:wavelength-selective} show how this could be done. Because of the physical separation between the leakage term and coronagraphic PSFs, a simple pick-off mirror can send the leakage light to a wavefront sensor. Using a wavefront sensor after the coronagraph has the advantage that the important non-common path aberrations can be removed. In addition, this increases flexibility as any free pupil plane in an existing instrument can be used for adding a vAPP. \\
The steepness of the transition of the band is determined by the number of liquid-crystal layers and their thickness. For Fig. \ref{fig:wavelength-selective} a 3-layered MTR was simulated with a total thickness of 20.7 micron. Going to thicker and more layers increases absorption of the liquid-crystals, the manufacturing process gets more difficult and the chance of dust contamination and costs increase. These disadvantages have to be weighed against the increase in bandwidth.
\section{Wavefront sensing applications of liquid-crystal technology}

The performance of high contrast imaging systems for ground based telescopes is very dependent on the correction of atmospheric turbulence. The aberrated wavefront needs to be precisely measured and corrected to reach the contrast levels required for imaging exoplanets. Here we discuss a few possibilities of applying liquid-crystal technology to improve wavefront sensing, using the properties that are enabled only by this technology. We focus on finding solutions for two difficulties in the field of wavefront sensing; non-common path aberrations and photon efficiency. \\
The best location for measuring non-common path aberrations is in the science focal plane. The phase diversity holograms introduced in Section \ref{sec:SCExAO} are an example of a focal plane wavefront sensor that can be used for measuring non-common path aberrations. This technique benefits from the extreme patterning and geometric phase. Similarly, a liquid-crystal version of the coronagraphic Modal Wavefront Sensor \cite{wilby2016designing,Wilby2016TheEnvironments} is an improvement over applying the phase with a spatial light modulator. Furthermore, the patterns required for focal-plane electric field sensing with pupil-plane holograms\cite{Por2016Focal-planeHolograms} can be easily manufactured with the liquid-crystal technology. All three focal plane wavefront sensors form a natural combination with the vAPP, operate in the same wavelength range as the coronagraph and can be implemented simultaneously without reducing the light in the coronagraphic PSFs by a significant amount.\\
Precise photon efficient wavefront sensing is possible with the Zernike wavefront sensor \cite{zernike1934diffraction,ZERNIKE1942974,Zernike1942PhaseI,Bloemhof2003PhaseOptics}. The Zernike wavefront sensor (ZWFS) applies a $\pi/2$ phase offset to the core of the PSF with a diameter of $1.06 \lambda/D$. The interference between the phase-shifted core and the rest of the PSF creates an intensity pattern in the consecutive pupil that is dependent on the phase of the incoming wavefront. This wavefront sensing technique has been demonstrated in the lab and on-sky \cite{vorontsov2001adaptive,Wallace2011Phase-shiftingSensorb,NDiaye2013CalibrationSensor,NDiaye2016CalibrationVLT/SPHERE}. It has been proven that the ZWFS is the most photon efficient wavefront sensor \cite{Guyon2005LimitsImaging,Paterson2008TowardsLimit}. The ZWFS however, suffers from chromaticity and it is both sensitive to amplitude and phase aberrations, where disentangling the two is not trivial. Here we propose the liquid-crystal version of the ZWFS, the vector-Zernike wavefront sensor (zWFS). The zWFS applies the phase of the normal ZWFS in the exactly the same way as a vAPP, with geometric phase. There are multiple advantages using geometric phase. The first one is that the applied phase is no longer dependent on the wavelength and the vZWFS can be applied for a larger bandwidth \cite{Bloemhof2014ApplicationOptics}. Still some chromaticity remains as the PSF scales with wavelength and the pattern does not. Secondly, the applied phase is $\pm \pi/2$ for opposite handedness of circular polarization. Wallace et al. (2011) \cite{ Wallace2011Phase-shiftingSensorb} do a similar thing with a phase-shifting ZWFS, however only one phase shift can be applied at the same time. Separating the opposite polarization states with a quarter-wave plate and a Wollaston prism allows to simultaneously measure two phase shifts ($\pm \pi/2$). The normalized difference of the two pupil measurements is directly proportional to the phase aberration and the sum of the two measurements is directly proportional to  amplitude aberrations. Therefore, the vector-Zernike wavefront sensor can simultaneously measure amplitude and phase aberrations with very simple operations, allowing for fast wavefront reconstruction. The dynamic range of the vZWFS is the same as the ZWFS, around 2 radians peak to valley. The vZWFS can best be used for precise measurements and correction of the residual wavefront after an adaptive optics system.  \\
When using a multi-step wavefront correction scheme, the first stage needs a wavefront sensor with large dynamic range that can deal with the large aberrations introduced by the atmosphere. A Shack-Hartmann wavefront sensor (SHWFS) and a (modulated) pyramid wavefront sensor have been widely used for this. Both the SHWFS and the pyramid WFS are not optimal as the resolution of the SHWFS and the dynamic range of the pyramid WFS is limited \cite{Guyon2005LimitsImaging,Ragazzoni1996PupilPrism}. A modulated pyramid WFS has increased dynamic range at a cost of sensitivity \cite{Burvall2006LinearitySensor.}. The generalised optical differentiation wavefront sensor (G-ODWFS) both has a large dynamic range and good sensitivity \cite{Haffert2016GeneralisedSensor}. The G-ODSWFS operates as a standard optical differentiation WFS (ODWFS), where an amplitude filter in the focal plane is used to measure the gradient of the wavefront \cite{Sprague1972QuantitativeObjects.}. For first lab tests, an SLM with crossed polarizers was used to create the amplitude filters. Because the G-ODWFS uses two filters per gradient direction with opposite amplitude, using liquid-crystal technology is a natural upgrade to enhance photon efficiency. A half-wave retarder with varying fast-axis orientation combined with a Wollaston prism creates exactly such an amplitude mask with opposite sign for each linear polarization output. The 100\% photon efficient G-ODWFS uses 2 Wollaston prisms and a liquid-crystal device \cite{Haffert2016GeneralisedSensor}. 

\section{Conclusion}
Writing extreme liquid-crystal patterns with the direct-write technology combined with self-aligning birefringent liquid-crystals allows for great flexibility in the creation of the vector Apodizing Phase Plate (vAPP) coronagraph. Any leakage term for a vAPP with a 360-degree dark zone can be decreased by up to two orders of magnitude by using a grating-vAPP in combination with a polarization grating with the same spatial frequency. The double-grating does not suffer from the large dispersion introduced by the grating-vAPP and therefore planet light can be spectroscopically analyzed when put into a fiber or integral field unit. A double-grating vAPP operating with an efficiency higher than 97\% from 2-5$\mu$m was installed on LBT. Observations of Regulus demonstrate the double-grating concept works. \\
Secondly, we introduce a grating mask for the vAPP that uses a polarization grating as an amplitude mask. Previously, a chromium mask was used to remove any light outside of the pupil. This adds an extra substrate with a chromium mask that has to be precisely aligned and glued on top of the vAPP. The grating-mask removes the need for such a mask by scattering the light outside the pupil to very high spatial frequencies. This light can be easily removed with a field stop. The concept has been successfully tested in the lab with the simplified WFIRST pupil. \\
Furthermore, we show a new grating-vAPP that will be installed in SCExAO on the Subaru telescope in the beginning of August 2017. The vAPP will operate with more than 98\% efficiency in J, H and K band and is able to feed the CHARIS instrument. The dark zone extends 2-11 $\lambda/D$ with a contrast of $10^{-5}$ across the three bands. Low-order focal plane wavefront sensing is possible due to the addition of phase diversity holograms. \\
Additionally, we report on wavelength-selective coronagraphs, with an optimized retardance function of wavelength such that they create a coronagraphic PSF at certain wavelenghts and transmit light unapodized at other wavelenghts. Wavefront sensors can be operated after wavelength-selective devices for unapodized light. This increases flexibility of upgrading existing instruments with a vAPP and adds the capability to do wavefront sensing of non-common path aberrations. \\
Lastly, we discuss several applications of liquid-crystal technology for wavefront sensing. Non-common path aberrations can be measured with multiple techniques using holograms created with geometric phase patterns and can be used in combination with a vAPP coronagraph. Liquid-crystal technology can also be used as an upgrade to existing wavefront sensing devices. The vector-Zernike WFS, can be used to simultaneously measure phase and amplitude aberrations when the opposite circular polarization states are separated on the detector by a quarter-wave plate and a Wollaston prism. Lastly, the generalized optical differentiation WFS can be manufactured with liquid-crystal technology, making it 100\% efficient. 

\acknowledgments 
The research of David S. Doelman and Frans Snik leading to these results has received funding from the European Research Council under ERC Starting Grant agreement 678194 (FALCONER)
\newpage
{\scriptsize
\bibliography{Mendeley}} 
\bibliographystyle{spiebib} 

\end{document}